\documentclass[conference]{IEEEtran}
\IEEEoverridecommandlockouts
\usepackage{cite}
\usepackage{amsmath,amssymb,amsfonts}
\usepackage{algorithmic}
\usepackage{graphicx}
\usepackage{algorithm,algorithmic}
\usepackage{textcomp}
\usepackage{xcolor}
\usepackage{bm}
\def\BibTeX{{\rm B\kern-.05em{\sc i\kern-.025em b}\kern-.08em
    T\kern-.1667em\lower.7ex\hbox{E}\kern-.125emX}}
 \newtheorem{lemma}{Lemma}

\newtheorem{theorem}{Theorem}

\begin{document}

\title{Acceleration of Cooperative Least Mean Square via Chebyshev Periodical 
Successive Over-Relaxation \\
\thanks{This work was partly supported by JSPS Grant-in-Aid for Scientific Research (B) 
Grant Number 19H02138 (TW) and 
for Early-Career Scientists Grant Number 19K14613 (ST).}
}

\author{
  \IEEEauthorblockN{Tadashi Wadayama\IEEEauthorrefmark{1} 
                and Satoshi Takabe\IEEEauthorrefmark{1}\IEEEauthorrefmark{2}
                }%\\
  \IEEEauthorblockA{\IEEEauthorrefmark{1}%
		Nagoya Institute of Technology,
		Gokiso, Nagoya, Aichi 466-8555, Japan, 
 		\{wadayama, s\_takabe\}@nitech.ac.jp} %\\
  \IEEEauthorblockA{\IEEEauthorrefmark{2}%
  		RIKEN Center for Advanced Intelligence Project,
  		Nihonbashi, Chuo-ku, Tokyo 103-0027, Japan
                }
} 
\maketitle

\begin{abstract}
A distributed algorithm for least mean square (LMS)
can be used in distributed signal estimation and 
in distributed training for multivariate regression models.
The convergence speed of an algorithm is a critical factor because 
a faster algorithm requires less communications overhead and
it results in a narrower network bandwidth.
The goal of this paper is to present that use of Chebyshev periodical 
successive over-relaxation (PSOR) can accelerate distributed LMS algorithms 
in a naturally manner. 
The basic idea of Chbyshev PSOR 
is to introduce 
index-dependent PSOR factors that control the spectral radius of a matrix governing the convergence 
behavior of the modified fixed-point iteration. Accelerations of convergence speed are 
empirically confirmed in a wide range of networks, such as 
known small graphs (e.g., Karate graph),  and random graphs, such as
Erd\"os-R\'enyi (ER) random graphs and  Barab\'asi-Albert random graphs.
\end{abstract}


\section{Introduction}

It is expected that a massive number of terminals will be connected in networks using beyond 5G/6G standards.
In such a situation, {\em cooperative signal processing} with neighboring nodes becomes 
more significant for enhancing the performance of signal processing regarding wireless communications.
For example, assume a case of a massive MIMO detection. If base stations are allowed to 
exchange certain information among the neighboring base stations, there are opportunities to 
improve the detection performance as if virtual multiple receive antennas were composed.
When a number of sensors are trying to learn a common multivariate regression  model  based on
own local data, a distributed algorithm for the least mean square (LMS) may be a natural choice as a learning strategy.

In the field of machine learning, {\em federated learning} \cite{Bonawitz}, which
is commonly implemented as
a distributed algorithm with a centralized parameter server,
is becoming a hot research topic.
Fully distributed algorithms such as the average consensus  algorithm \cite{Xiao}, 
which has no centralized server,  
have several advantages over the centralized distributed algorithms.
One of the advantages is robustness,  that is,  even when some of nodes stop operating
because of a malfunction or dead battery, 
a fully distributed algorithm often can keep working. Another merit of fully distributed algorithms
is that it can balance signal traffics over a network. A centralized distributed algorithm 
often creates unbalanced network traffics,  where the edges connected to the centralized server 
needs to accommodate the largest amount of traffics,  and traffics on other edges in the network
is much smaller than the traffic at the centralized server.

Diffusion LMS \cite{Cattivelli, Sayed, Lopes} is a notable example of 
fully distributed estimation algorithm.
The core of the diffusion LMS consists of two steps. The first step can be 
seen as a local LMS estimation and the second step is to diffuse the local estimations to 
neighboring nodes. All the agents in the network repeatedly execute these two steps and 
eventually all the agent states converges to the global solution.  Sayed et al.  \cite{Sayed}
reported an analysis of the convergence rate of diffusion LMS,  and showed advantages 
both in stability and convergence rate. An acceleration method for Diffusion LMS
based on belief propagation was discussed in \cite{nakai}.
A consensus-based distributed LMS algorithm was presented by Schizas et al. \cite{Schizas}. 
Their derivation of the algorithm introduced auxiliary local variables for each agent 
and provided a global objective function that naturally fits the problem setting. The minimization 
problem for the global objective function can be cast as a convex constrained minimization 
problem. The proposed algorithm in \cite{Schizas} is naturally derived from the ADMM formulation for solving 
the convex problem.
Decentralized baseband signal processing of MIMO detection was discussed by Li et al. \cite{Li}. MIMO signal detection 
is closely related to LMS problems. Decentralized baseband signal processing appears promising 
for reducing the prohibitive complexity of handling baseband signal processing with a massive number of antennas.

The authors proposed a method to accelerate the convergence of a fixed-point iteration in \cite{takabe20}.
The acceleration method is called {\em Chebyshev periodical successive over-relaxation (Chbyshev PSOR)}, 
and it  is applicable 
both to linear and non-linear fixed-point iterations. 
The basic idea of Chbyshev PSOR 
is to introduce 
index-dependent PSOR factors that control the spectral radius of a matrix governing the convergence 
behavior of the modified fixed-point iteration. The name of the method is named after the Chebyshev polynomials
that are used for determining the PSOR factors. In \cite{takabe20}, 
it is shown that many fixed-point iterations,
such as the Jacobi method for solving linear equations  are successfully accelerated.

It would be very natural to use Chebyshev PSOR for accelerating a distributed LMS algorithm
because most of a fully distributed LMS algorithm can be regarded as linear fixed-point iterations.
Acceleration of a fully distributed LMS algorithm in convergence seems an appropriate  problem to pursue because
a fast algorithm generates less signal traffics over a network and it reduces computational complexity 
for each nodes.

The goal of this paper is to show that the use of Chebyshev-PSOR can accelerate distributed LMS algorithms 
in a naturally manner. We thus place our main focus on how to accelerate a fully distributed LMS algorithm,  which 
is referred to as {\em cooperative LMS}. The cooperative LMS which is derived from the global objective function 
via the use of a proximal gradient method \cite{Prox}. 
Our intension is not to develop the  fastest algorithm but to {\em present  the principle for accelerating 
the convergence of a  distributed LMS algorithm}.  Chebyshev PSOR can be applied to another distributed LMS
algorithms but we restrict our attention to the cooperative LMS to keep the discussion focused.
Cooperative LMS is closely related to the diffusion LMS  \cite{Cattivelli, Sayed, Lopes} 
and other distributed LMS algorithms \cite{Schizas}. It is expected that the results shown in this paper 
is applicable to these algorithms as well.

\section{Preliminaries}

%This section provides background, basic problem setup and notation required for the paper.

\subsection{Notation} 
The range of integers from $1$ to $N$ is represented as $[N]$. The closed real 
interval from $a$ to $b$ is denoted by $[a,b]$ and the open interval is denoted by $(a, b)$.
Let $\bm A$ be an $n \times n$ real symmetric matrix.  The notation $\lambda_{min}(\bm A)$ and $\lambda_{max}(\bm A)$
denote the minimum and maximum eigenvalues of $\bm A$, respectively. The notation $\rho(\bm A)$ indicates 
the spectral radius of $\bm A$. The matrix $\bm I_n$ means the identity matrix of size $n \times n$.
If the size is evident from the context, the identity matrix  is simply denoted by $\bm I$.
The operator $\otimes$ represents the Kronecker product.

\subsection{Problem setup}

Let $G :=(V, E)$ be an undirected connected graph representing a network of agents.
Assume that an agent $k \in V := \{1,2,\ldots, K\}$ can communicate with 
its neighboring agents in  ${\cal N}(k) := \{j \mid (k, j) \in E \}$. 
Each agent has own observation vector $\bm y_k \in \mathbb{R}^m (m < N)$ which is given by
$
	\bm y_k = \bm H_k \bm x^0+ \bm w_k
$
where $\bm x^0 \in \mathbb{R}^N$ is a parameter vector unknown to all agents 
and $\bm H_k \in \mathbb{R}^{m \times N}$ is a real matrix known to agent $k$.
The additive term $\bm w_k$ represents i.i.d. Gaussian noise vector where 
each component follows Gaussian distribution with zero mean and variance $\sigma^2$.
To estimate the hidden global parameter $\bm x^0$, we can use the LMS estimation defined as
$
	\bm x^* := \text{argmin}_{\bm x \in \mathbb{R}^N} \sum_{k = 1}^K (1/2) \|\bm y_k -  \bm H_k\bm x \|^2.
$
In the distributed environment defined above, it is natural to employ a gradient descent method 
to solve the LMS  problem. 

\subsection{Brief review of Chebyshev PSOR}

In this subsection, we will briefly review 
some basic facts regarding Chebyshev PSOR
according to \cite{takabe20}.

Let us consider the following linear 
fixed-point iteration:
\begin{equation}
\bm x^{(k+1)} := \bm A \bm x^{(k)}, \ k = 0, 1, 2, \ldots 
\end{equation}
where $\bm A \in \mathbb{R}^{n \times n}$
and $\bm x^{(t)} \in \mathbb{R}^n$ for $k = 0,1,2,\ldots$ as an example.
The method in \cite{takabe20} handles more general fixed-point iterations, such as 
$\bm x^{(k+1)} := f(\bm x^{(k)})$ but we here restrict our attention to the simplest case 
required for the following discussion. If the spectral radius of $\bm A$ satisfies 
$\rho(\bm A) < 1$, the linear fixed-point iteration converges to the fixed point $\bm x^* = \bm 0$. 

Successive over-relaxation (SOR) is a well-known method 
for accelerating this fixed point iteration 
with the modified fixed point iteration: 
\begin{align} \label{PSORitr} 
\bm x^{(k+1)} &:= \bm x^{(k)} 
+ \omega_k\left( \bm A \bm x^{(k)} - \bm x^{(k)}  \right), 
\end{align}
where $\omega_k (k = 0, 1, \ldots) $ is 
a positive real number called a SOR factor.
In this paper, we will use {\em periodical SOR (PSOR) factors} $\{\omega_k \}_{k=0}^{T-1}$ satisfying 
$
\omega_{\ell T + j} = \omega_j \quad (\ell = 0,1,2,\ldots,\  j = 0, 1, 2, \ldots, T-1),
$
where $T$ is a positive integer called the {\em period} of the PSOR factors.
SOR using PSOR factors is referred to as a {\em PSOR}.
The PSOR iteration (\ref{PSORitr}) can be rewritten in a linear update form:
\begin{equation}
\bm x^{(k+1)} := (\bm I - \omega_k \bm B ) \bm x^{(k)},	
\end{equation}
where the matrix $\bm B$ is defined by $\bm B := \bm I - \bm A$. By using the periodicity of $\omega_k$,
we immediately have an update equation for every $T$ iterations as 
\begin{align} \label{PSORitr2}
\bm x^{((\ell+1)T)} &= \left[\prod_{k = 0}^{T-1}(\bm I - \omega_k \bm B ) \right] \bm x^{(\ell T)}\\
&= \bm U(\{\omega_k\}_{k=0}^{T-1}) \bm x^{(\ell T)},
\end{align}
where $\bm U(\{\omega_k\}_{k=0}^{T-1}) := \prod_{k = 0}^{T-1}(\bm I - \omega_k \bm B )$.
From this equation,  we can observe that the dynamics of the linear update (\ref{PSORitr2}) is governed by 
the eigenvalues of  $\bm U(\{\omega_k\}_{k=0}^{T-1})$ and we can control the PSOR coefficients
$\{\omega_k\}_{k=0}^{T-1}$  to accelerate the convergence of  (\ref{PSORitr2}).
To find a suboptimal set of $\omega_k$,  the polynomial defined by
\begin{align}
\beta(\lambda; \{\omega_k\}_{k=0}^{T-1}) = 	
\left[\prod_{k = 0}^{T-1}(1 - \omega_k \lambda ) \right] 
\end{align}
is a useful tool. Let $\lambda_1 \le  \lambda_2 \le  \cdots \le \lambda_n$ be the eigenvalues of $\bm B$. It is known that
the eigenvalues of $\bm U(\{\omega_k\}_{k=0}^{T-1})$ can be represented as
\begin{align} \nonumber
	\{ \beta(\lambda_1; \{\omega_k\}_{k=0}^{T-1}), \beta(\lambda_2; \{\omega_k\}_{k=0}^{T-1})\ldots, \beta(\lambda_n; \{\omega_k\}_{k=0}^{T-1}) \}.
\end{align}
This fact inspires us to choose a polynomial with small absolute value in the range $[\lambda_1, \lambda_n]$
to determine the PSOR factors $\{\omega_k\}_{k=0}^{T-1}$ because smaller absolute values of eigenvalues  $\bm U(\{\omega_k\}_{k=0}^{T-1})$ lead to faster convergence.

%If $\bm U(\{\omega_k\}_{k=0}^{T-1})$ is diagonalizable, 	$\bm U(\{\omega_k\}_{k=0}^{T-1})$ can be represented as
%$
%	\bm U(\{\omega_k\}_{k=0}^{T-1}) = \bm P \bm D \bm P^{-1}, 
%$
%where $\bm D$ is the diagonal matrix with eigenvalues of $\bm U(\{\omega_k\}_{k=0}^{T-1})$ as the diagonal elements
%and $\bm P$ is a matrix composed form eigenvectors as column vectors. 
%Assume also that $ \bm x^{(\ell T)} = \bm P \bm v$ for a vector $\bm v = (v_1,\ldots, v_n)^T \in \mathbb{R}^n$.
%Under this situation,  $\|\bm x^{((\ell+n)T)} \|$ can be upper bounded as
%\begin{align}
%\|\bm x^{((\ell+m)T)} \| 
%%&= \left\| \left[\prod_{k = 0}^{T-1}(\bm I_n - \omega_k \bm B ) \right]^m \bm x^{(\ell T)} \right\| \\
%%&= \| \bm P \bm D^m \bm P^{-1} \bm P \bm v  \| \\
%%&= \|\bm P \bm D^m \bm v \| \\
%%&\le \|\bm P \| \| \bm D^m \bm v\| \\ \label{lambda_power}
%&\le \|\bm P \|   \left|\sum_{i=1}^n \lambda_i^m v_i \right|,
%\end{align}
%where $\lambda_i (i \in [n])$ is an eigenvalue of $\bm U(\{\omega_k\}_{k=0}^{T-1})$.
%From this inequality,  it is evident that the choice $\{\omega_k\}_{k=0}^{T-1}$ is crucial to 
%decrease the absolute value of $\lambda_i$ for achieving faster convergence.

\begin{figure}
\begin{center}
\includegraphics[scale=0.4]{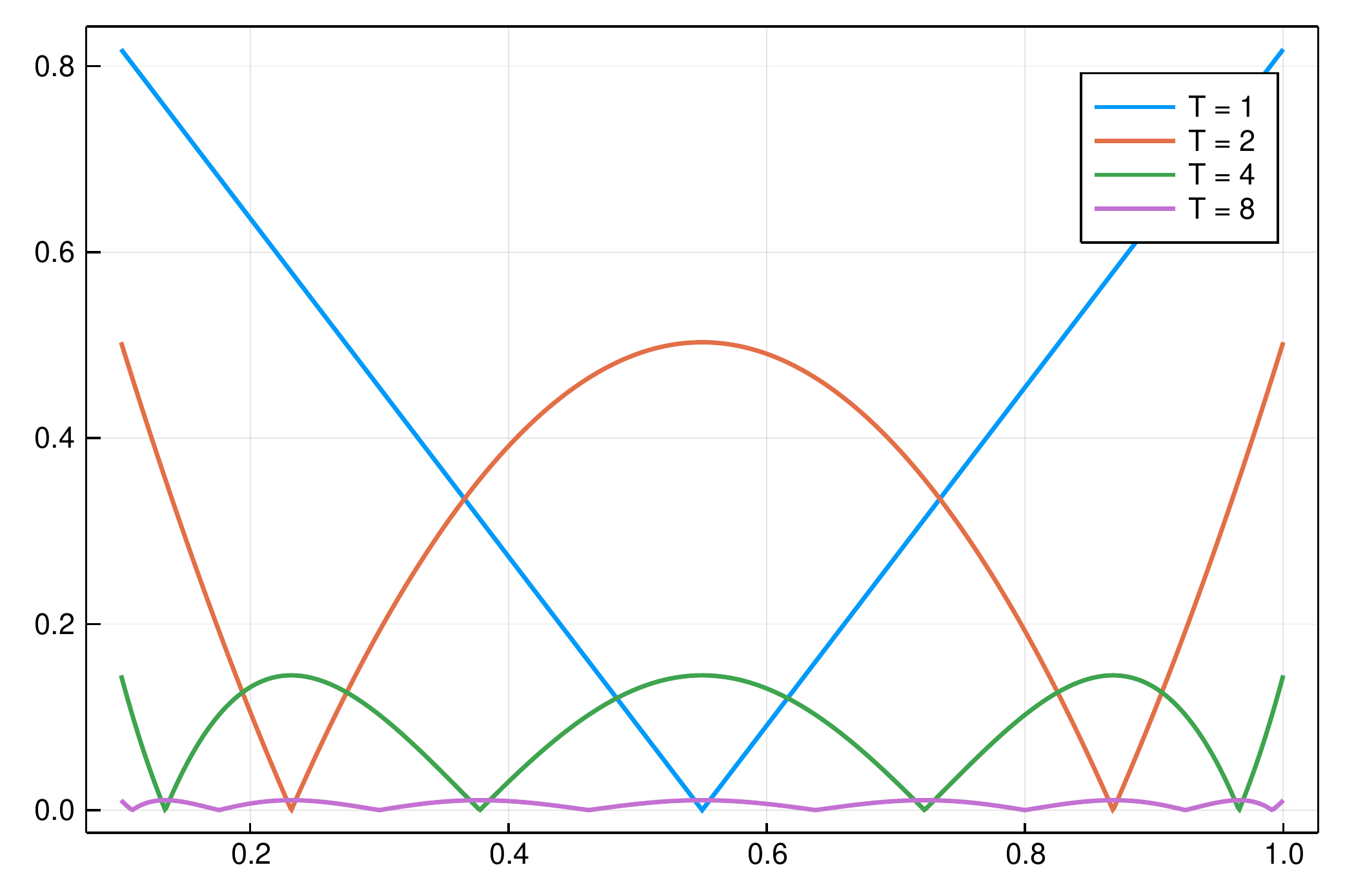}
\end{center}
\caption{Plot of the absolute values of $\beta(\lambda; \{\omega_k^{ch}\}_{k=0}^{T-1})$ for $T = 1,2,4, 8$.
The hyperparameters are set to $a = 0.1$ and $b = 1.0$.} 
\label{beta_plot}
\end{figure}

The paper \cite{takabe20} introduced an {\em affine translate of the Chebyshev polynomial} having the desired properties.
We define the {\em Chebyshev PSOR factors} $\{\omega_k^{ch}\}_{k=0}^{T-1}$ for the range $[a,b]$ by
\begin{equation} \label{SORfactors}
\omega_k^{ch}
    :=\left[ \frac{b+a}{2} + \frac{b-a}{2} \cos \left(\frac{2k+1}{2T} \pi \right) \right]^{-1},
\end{equation}
which are reciprocals of the roots of an affine translated Chebyshev polynomial.
It is shown in \cite{takabe20} that the polynomial defined by the Chebyshev PSOR factors 
$\beta(\lambda; \{\omega_k^{ch}\}_{k=0}^{T-1})$ has tightly bounded absolute values in the range $[a, b]$.
Figure \ref{beta_plot} displays the absolute value of $\beta(\lambda; \{\omega_k^{ch}\}_{k=0}^{T-1})$ 
for $T = 1,2,4, 8$ under the setting $a = 0.1$ and $b = 1.0$.
We can readily  confirm that the absolute values of the function in the range $[0.1, 1.0]$ are tightly bounded.
\section{Derivation of Cooperative LMS Algorithm}

%In the following, we will discuss a distributed algorithm to solve the LMS problem.

\subsection{Distributed LMS problem as a regularization problem}

Let $\bm x_k (k \in [K])$ be a state vector corresponding to the agent $k$, which 
represents a tentative estimate of $\bm x_0$.
We here introduce 
a global loss function as 
\begin{align} \label{objfunc}
	\ell(\bm\chi) := \frac 1 2 \sum_{k \in [K]}\|\bm y_k - \bm H_k \bm x_k\|^2 +  \eta \bm\chi^T   (\bm L \otimes \bm I_N) \bm\chi,
\end{align}
where the global state vector $\bm\chi$ is defined by
\begin{equation}
	\bm\chi := \left( 
	\begin{array}{c}
      \bm x_1  \\
%      \bm x_2  \\
	\vdots \\
      \bm x_K  \\
	\end{array}
	\right).
\end{equation}
The matrix $\bm L := \bm D - \bm A$ is the graph Laplacian of $G$ where $\bm A$ is the 
adjacency matrix of $G$ and $\bm D$ is the degree matrix of $G$,  that is, 
the $(i, i)$ element of the diagonal matrix $\bm D$ is the degree of node $i$.
%The notable difference between the original LMS loss and 
%the loss $\ell(\bm\chi)$  is that $\ell(\bm\chi)$ contains agent-dependent states $\bm x_k$ but the 
%original one only contains the global state $\bm x$.
It should be noted that 
\begin{equation}
	\bm\chi^T    (\bm L \otimes \bm I_N)  \bm\chi = \frac 1 2 \sum_{(i,j) \in E} \|\bm x_i - \bm x_j \|^2	
\end{equation}
holds,  and it can be considered as a regularization term enforcing the proximity of neighboring agent states
$\bm x_i, \bm x_j ((i, j) \in E)$.
It is important to realize that the original LMS estimation and the minimizer 
$
	\bm\chi^* := \text{argmin}_{\bm\chi \in \mathbb{R}^{KN}} \ell(\bm\chi)
$
are closely related but their solution may not be the same. The LMS estimation defined by 
the global loss function  is denote by the {\em cooperative LMS estimation}.
The problem for minimizing the global loss function is a quadratic problem and it is 
strictly convex if certain conditions are satisfied.
In such a case,  the minimization problem for $\ell(\bm\chi)$ has a unique minimizer.

To solve the minimization problem regarding cooperative LMS,
we will employ the proximal gradient method \cite{Prox}.
The gradient descent step is simply given by the following update rule:
\begin{align} \label{grad_step}
\bm\chi^{(t+1)} := \bm\chi^{(t)} - \mu \nabla_{\bm\chi} \frac 1 2 \sum_{k \in [K]}\|\bm y_k - \bm H_k \bm x_k\|^2.
\end{align}
The proximal step can be approximated with a gradient descent step for the quadratic form 
$\eta \bm\chi^T   (\bm L \otimes \bm I_N) \bm\chi$, which can be given by
$
\bm\chi^{(t+1)} 
%&= \bm\chi^{(t)} - \eta \bm L \otimes \bm I_N \bm\chi^{(t)} \\
:= (\bm I - \eta \bm L \otimes \bm I_N) \bm\chi^{(t)}.
$
In the following subsections, we will discuss both steps in detail.

\subsection{Gradient descent step}

Assume that $\bm x_k^{(t)}$ is the state of agent $k$ at discrete time index $t$.
The gradient step (\ref{grad_step}) can be executed in parallel as 
\begin{eqnarray}
	\bm x_k^{(t+1)} 
	&:=& \bm x_k^{(t)} + \mu \bm H_k^T(\bm y_k - \bm H_k \bm x_k^{(t)}) \\
	&=& (\bm I - \mu \bm H_k^T \bm H_k) \bm x_k^{(t)} + \mu \bm H_k^T \bm y_k \\
	&=& \bm A_k \bm x_k^{(t)}  + \bm b_k,
\end{eqnarray}
where $\bm A_k := \bm I - \mu \bm H_k^T \bm H_k$ and  $\bm b_k = \mu \bm H_k^T \bm y_k$.
The global state vector $\bm\chi^{(t)}$ at time index $t$ and the offset vector $\bm\beta$ are defined by
\begin{equation}
	\bm\chi^{(t)} := \left( 
	\begin{array}{c}
      \bm x_1^{(t)}  \\
      \bm x_2^{(t)}  \\
	\vdots \\
      \bm x_K^{(t)}  \\
	\end{array}
	\right), \quad
	\bm\beta := \left( 
	\begin{array}{c}
      \bm b_1 \\
      \bm b_2  \\
	\vdots \\
      \bm b_K  \\
	\end{array}
	\right).
\end{equation}
Let $\bm D \in \mathbb{R}^{NK \times NK}$ be a block diagonal matrix consisting of $\bm A_1,\ldots, \bm A_K$ as diagonal block matrices: 
\begin{equation}
	\bm D := 
	\left(
	\begin{array}{cccc}
	\bm A_1 &               &             &      \\
	              & \bm A_2 &             &       \\
	              &               & \ddots  &       \\
                  &            	&             & \bm A _k  \\
	\end{array}
	\right).
\end{equation}
From these notation, we can compactly represent the gradient descent step  as
$
	\bm\chi^{(t+1)}  := \bm D \bm\chi^{(t)} + \bm\beta. 
$

\subsection{Average consensus protocol as proximal step}

Next, we consider an implementation of the proximal step.
We here introduce the simplest average consensus scheme based on the update equation:
\begin{equation} \label{consensus_itr}
\bm x_k^{(t+1)} := \bm x_k^{(t)} + \eta \sum_{j \in {\cal N}(k)} (\bm x_j^{(t)} - \bm x_k^{(t)}),
\end{equation}
where $\eta$ is a positive real number. If the parameter $\eta$ is appropriately determined,
the above iterations eventually converge to the average of the initial vectors. The process 
is known as {\em average consensus}.
A careful observation reveals 
that the consensus iteration defined by (\ref{consensus_itr}) can be represented by
\begin{equation} \label{consensus_global}
		\bm\chi^{(t+1)}  := (\bm I - \eta \bm L \otimes \bm I_N) \bm\chi^{(t)},
\end{equation}
which shows the equivalence between the proximal step defined above
and the average	 consensus protocol (\ref{consensus_itr}).

\section{Properties of Cooperative LMS Algorithm}

In the previous section,  we saw that 
the gradient descent step can be executed perfectly in parallel 
and that
the proximal step can be executed with the average consensus protocol which only requires
neighboring interactions between agents.
In this section, we will study {\em cooperative LMS} which is a realization of
the proximal gradient method.

\subsection{Implementation of cooperative LMS algorithm}

In this paper, we deal with the simple distributed LMS defined in Alg. \ref{dist_LMS}, 
which can solve  the cooperative LMS problem.
This algorithm is closely related  to diffusion LMS  \cite{Cattivelli, Sayed, Lopes}
 and consensus-based learning algorithms \cite{Schizas}.

\begin{algorithm}
 \caption{Cooperative LMS}
 \label{dist_LMS}
 \begin{algorithmic}[1]
 \renewcommand{\algorithmicrequire}{\textbf{Input:}}
 \renewcommand{\algorithmicensure}{\textbf{Output:}}
% \REQUIRE in
% \ENSURE  out
  \STATE For each $k \in [K]$, set  $\bm x^{(0)} := \bm 0$
  \FOR {$t : = 0$ to $L-1$}
  \FOR {$k := 1$ to $K$}
  \STATE \bm $\bm u_k^{(t)} 
  :=\bm x_k^{(t)} + \mu \bm H_k^T(\bm y_k - \bm H_k \bm x_k^{(t)})$ 
	\STATE $\bm x_k^{(t+1)} := \bm u_k^{(t)} + \eta \sum_{j \in {\cal N}(k)} (\bm u_j^{(t)} - \bm u_k^{(t)})$
	\ENDFOR
  \ENDFOR
 \RETURN $\bm x_1^{(L)}, \bm x_2^{(L)}, \ldots, \bm x_K^{(L)}$
 \end{algorithmic} 
 \end{algorithm}

For the following analysis,  it is desirable to have an equivalent algorithm that is based on 
the update equation for $\bm\chi^{(t)}$.
Combining the two update rules,  specifically,   the gradient step  and the proximal step, we immediately 
have the following global update rule corresponding to  Alg. \ref{dist_LMS}:
\begin{eqnarray}\label{fixedpoint}
\bm\chi^{(t+1)}  &:=& (\bm I - \eta \bm L \otimes \bm I_N) ( \bm D \bm\chi^{(t)} + \bm\beta) \\ \nonumber
&=&  \bm Q \bm\chi^{(t)} + (\bm I - \eta \bm L \otimes \bm I_N) \bm\beta,
\end{eqnarray}
where
$
	\bm Q := (\bm I - \eta \bm L \otimes \bm I_N) \bm D. 
$
An equivalent algorithm shown in Alg. \ref{equiv_alg} includes an affine fixed point iteration 
whose error dynamics of the system is governed by the matrix: 
$
	\bm Q := (\bm I_{K N} - \eta \bm L \otimes \bm I_N) \bm D. 
$

\begin{algorithm}
 \caption{Cooperative LMS (equivalent form)}
 \label{equiv_alg}
 \begin{algorithmic}[1]
 \renewcommand{\algorithmicrequire}{\textbf{Input:}}
 \renewcommand{\algorithmicensure}{\textbf{Output:}}
  \STATE $\bm\chi^{(0)} = \bm 0$
  \FOR {$t = 0$ to $L-1$}
  \STATE $\bm\chi^{(t+1)} = \bm Q \bm\chi^{(t)}  + (\bm I_{K N} - \eta \bm L \otimes \bm I_N) \bm\beta$
  \ENDFOR
 \RETURN $\bm\chi^{(L)}$ 
 \end{algorithmic} 
 \end{algorithm}

Assume that the affine fixed-point iteration (\ref{fixedpoint}) has a fixed point 
$\bm\chi^*$ that satisfies 
\begin{align} 
\bm\chi^{*}  
= (\bm I - \eta \bm L \otimes \bm I_N) \bm D \bm\chi^{*}
  + (\bm I - \eta \bm L \otimes \bm I_N) \bm\beta. 
\end{align}
By subtracting the above equation from (\ref{fixedpoint}), we immediately obtain
a fixed-point iteration representing the evolution of error:
\begin{align} \label{error_evolution}
\bm\chi^{(t+1)} - \bm\chi^{*}  
= \bm Q (\bm\chi^{(t)} - \bm\chi^{*}).
\end{align}
From this error evolution equation, we can prove the following inequality 
which indicates linear convergence of the cooperative LMS algorithm 
to the global minimum if $\lambda_{max}(\bm D) < 1$.
\begin{lemma}
If $\bm Q$ is positive definite,  the following inequality holds:
\begin{align}\label{linear_convergence}
\frac{\|\bm\chi^{(t+1)} - \bm\chi^{*} \|}{\|\bm\chi^{(t)} - \bm\chi^{*}\|}
\le  \lambda_{max}(\bm D).
\end{align}
\end{lemma}
(Proof) By taking the norm of both sides of (\ref{error_evolution}), we have
\begin{align}
\| \bm\chi^{(t+1)} - \bm\chi^{*}  \|
&= \| \bm Q (\bm\chi^{(t)} - \bm\chi^{*}) \|  \\
&\le \|(\bm I_{K N} - \eta \bm L \otimes \bm I_N)\| \|\bm D\| \|\bm\chi^{(t)} - \bm\chi^{*}\|\\
& = \|\bm D\| \|\bm\chi^{(t)} - \bm\chi^{*}\| \\
& = \rho(\bm D) \|\bm\chi^{(t)} - \bm\chi^{*}\|  \\
& = \lambda_{max}(\bm D) \|\bm\chi^{(t)} - \bm\chi^{*}\|,
\end{align}
where the second inequality uses the fact $\|(\bm I_{K N} - \eta \bm L \otimes \bm I_N)\| =1$.
\hfill \fbox{}

\subsection{Smallest and largest eigenvalues of $\bm Q$}

The matrix $\bm Q$ should be positive definite 
so that the objective function (\ref{objfunc}) becomes strictly convex.
Furthermore, the eigenvalues of $\bm Q$ are of critical importance because they determines 
the convergence behavior of Chebyshev PSOR. 
In the following, we discuss the positive definiteness of $\bm Q$.

The following two lemmas will be basis of the positive definiteness of $\bm Q$.
\begin{lemma}
If $\eta < 1/\lambda_{max}(\bm L)$, then 	$\bm I_{K N} - \eta \bm L \otimes \bm I_N$ is positive definite.
\end{lemma}
(Proof)
Let $\bm A, \bm B \in \mathbb{R}^{n \times n}$. It is known that the set of eigenvalues 
of $\bm A \otimes \bm B$ is given by
$\{\lambda_A \lambda_B  \mid \lambda_A \in \Lambda_A, \lambda_B \in \Lambda_B \}$ where
$\Lambda_A$ and $\Lambda_B$ is the set of eigenvalues of $\bm A$ and $\bm B$, respectively.
The eigenvalues of $\eta \bm L \otimes \bm I_N$ is thus in the range 
$[\eta \lambda_{min}(\bm L),  \eta \lambda_{max}(\bm L)]$.
Then, the minimum eigenvalue of $\bm I_{K N} - \eta \bm L \otimes \bm I_N$ becomes 
$1 - \eta \lambda_{max}(\bm L)$. Due to the assumption $\eta < 1/\lambda_{max}(\bm L)$, 
we have $\lambda_{min}(\bm I_{K N} - \eta \bm L \otimes \bm I_N) > 0$. \hfill \fbox{}

\begin{lemma}
	If  $\lambda_{max}(\bm H_k^T \bm H_k) < 1/\mu$ for all $k \in [K]$, then $\bm D$ is positive definite.
\end{lemma}
(Proof) The claim is equivalent to the positive definiteness of all the matrices $\bm I - \mu \bm H_k^T \bm H_k$
under the condition $\lambda_{max}(\bm H_k^T \bm H_k) < 1/\mu$ for all $k \in [K]$. The minimum eigenvalue 
can be evaluated as 
\begin{equation}
	\lambda_{min}(\bm I - \mu \bm H_k^T \bm H_k) = 1 - \mu \lambda_{max}(\bm H_k^T \bm H_k).
\end{equation}
When the condition is met,  we immediately have $\lambda_{min}(\bm I - \mu \bm H_k^T \bm H_k) > 0$ 
for any $k \in [K]$. \hfill\fbox{}

The following theorem clarifies when $\bm Q$ is guaranteed to be positive definite.
\begin{theorem}
	If $\eta < 1/\lambda_{max}(\bm L)$ and $\lambda_{max}(\bm H_k^T \bm H_k) < 1/\mu$ for all $k \in [K]$,
	$\bm {Q}$ is positive definite and all the eigenvalues of $\bm Q$ are real.
\end{theorem}
(Proof) Assume that two Hermitian matrices $\bm X, \bm Y$ are both positive definite. Then, $\bm X \bm Y$ is also positive definite and all the eigenvalues of $\bm X \bm Y$ are real. Under the given condition, 
$\bm I_{K N} - \eta \bm L \otimes \bm I_N$ and $\bm D$ are both hermitian and positive definite
from above lemmas. The claim of the theorem follows from the above product property. \hfill\fbox{}

We next examine the largest eigenvalue of $\bm Q$. If $\bm Q$ is positive definite,
the largest eigenvalue of $\bm Q$ coincides with the spectral radius of $\bm Q$. 

\begin{theorem} \label{eigen_range}
If $\eta < 1/\lambda_{max}(\bm L)$ and $\lambda_{max}(\bm H_k^T \bm H_k) < 1/\mu$ for all $k \in [K]$,
any eigenvalue of $\bm Q$ is in the range $(0, 1)$.
\end{theorem}
(Proof)
We can upper bound  the largest eigenvalue of $\bm Q$ in the following way:
\begin{align}
	\lambda_{max}(\bm Q) & = \rho(\bm Q)   \\
	&\le \|\bm Q\| \\
	&\le \|(\bm I_{K N} - \eta \bm L \otimes \bm I_N)\| \|\bm D\| \\
	&= \max_{k=1}^K \lambda_{max}(\bm I_N - \mu \bm H_k^T \bm H_k) \\
	&= 1 - \mu \min_{k=1}^K \lambda_{min}(\bm H_k^T \bm H_k)\\
	&< 1,
\end{align}
where the first inequality is due to the norm upper bound for the spectral radius. 
The second inequality is based on the sub-additivity of the operator norm.
Since the laplacian $\bm L$ has zero eigenvalue, we have  
$ \|(\bm I_{K N} - \eta \bm L \otimes \bm I_N)\| = \rho(\bm I_{K N} - \eta \bm L \otimes \bm I_N) = 1$.
Combining the claim of Theorem 1, we have the claim of this theorem.
\hfill\fbox{}

From the proof of this theorem, we get to know that 
\begin{align}
\lambda_{max}(\bm D) < 1
\end{align}
if $\eta < 1/\lambda_{max}(\bm L)$ and $\lambda_{max}(\bm H_k^T \bm H_k) < 1/\mu$ for all $k \in [K]$.
In order to have the lowest $\lambda_{max}(\bm D)$ to get 
the fastest convergence rate in (\ref{linear_convergence}),
an optimal choice of the parameters would be 
\begin{align} \label{eta_rule}
	\eta &= \frac{1-\epsilon}{\lambda_{max}(\bm L)}, \\ \label{mu_rule}
	\mu  &= \frac{1-\epsilon}{\max_{k =1}^K\lambda_{max}(\bm H_k^T \bm H_k)}
\end{align}
where $\epsilon$ is a small positive real number.
It is easy to confirm that these parameter setting 
satisfies the positive definiteness conditions on $\eta$ and $\mu$.

\subsection{Validation of choice of $\eta$ and $\mu$}
\label{karate_experiment}
In this section,  we will provide experimental validation for the choice of 
$\eta$ and $\mu$ given by (\ref{eta_rule}) and (\ref{mu_rule}) where 
we will confirm whether these parameter setting practically provides 
fast convergence or not.

The experimental conditions are summarized as follows.
We used Karate graph with $K = 34$.
The dimension of the $\bm x_0$ is set to $N = 3$ 
where each element in $\bm x_0$ follows ${\cal N}(0,1)$.
Each element in $H_k $ follows ${\cal N}(0,1)$ where 
$H_k \in \mathbb{R}^{2 \times 3}$, i.e., $m = 2$.
The standard deviation of the observation noises is set to $\sigma = 0.1$.
The parameter $\epsilon$ used in (\ref{eta_rule}) and (\ref{mu_rule}) is set to $0.05$.
In order to estimate the expectation, we run 100-trials.

\begin{figure}
\begin{center}
\includegraphics[scale=0.45]{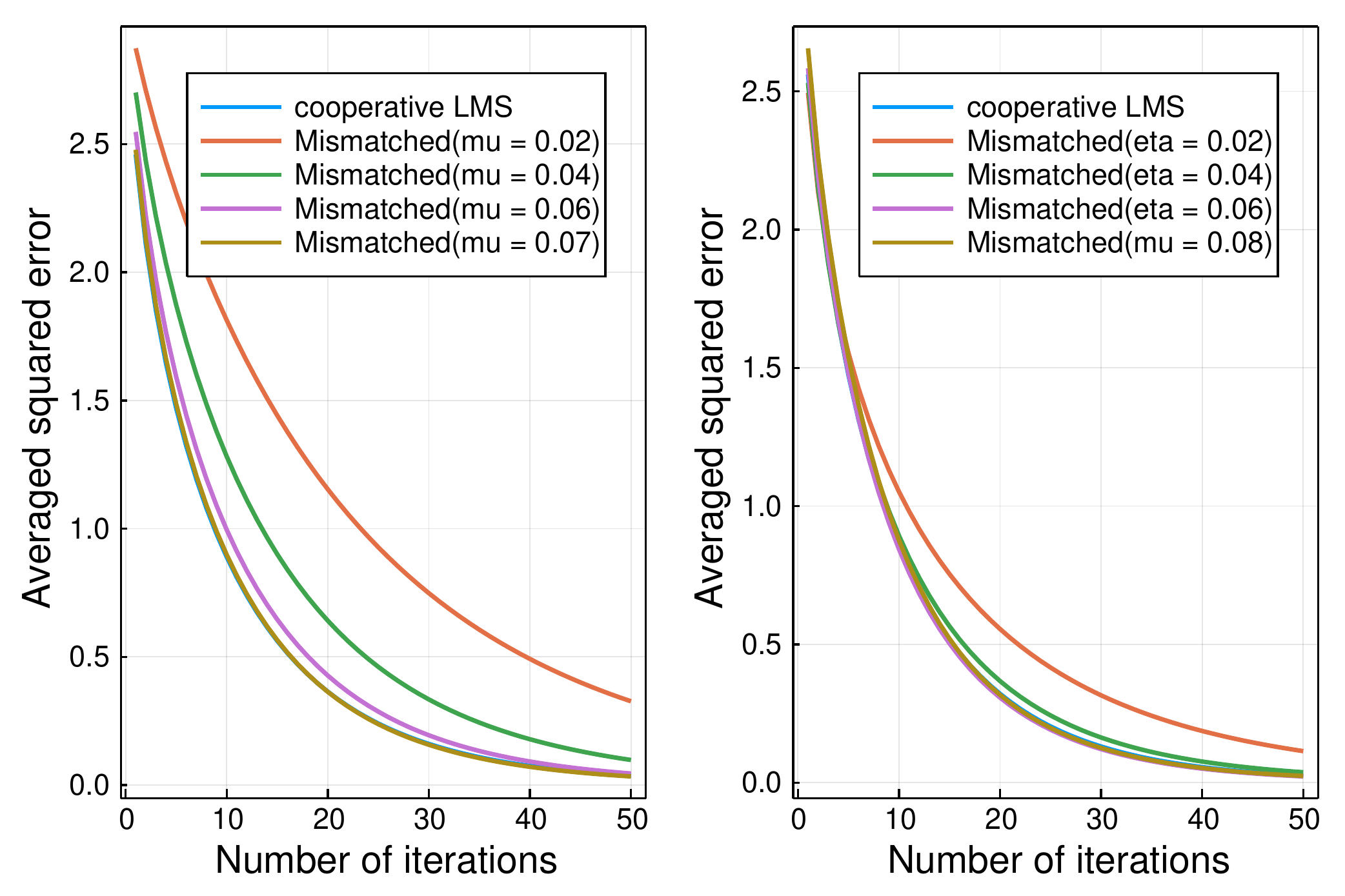}
\end{center}
\caption{Averaged squared errors of mismatched LMS: (left) fixed $\eta$ determined by (\ref{eta_rule}), (right) fixed $\mu$ determined by  (\ref{mu_rule}).} 
\label{mismatched}
\end{figure}

In the first experiment, while we fix the parameter $\eta$ as the value determined by (\ref{eta_rule}),
we use several {\em mismatched values} of $\mu$ in the cooperative LMS defined by Alg. \ref{dist_LMS}.
Namely,  the ASE performance of mismatched LMS's are examined here.
Figure \ref{mismatched} (left) presents the ASEs as the function of the number of iterations.
The ASE of the mismatched LMS with $\mu \in \{0.02, 0.04, 0.06, 0.07\}$ are presented.
As a baseline for comparison,  ASE of the cooperative LMS 
with the parameter determined by  (\ref{eta_rule}) and (\ref{mu_rule}) is also included in the figure.
In this case, the average value of $\mu$ given by (\ref{mu_rule}) is $0.0741$.
We can immediately observe that the convergence becomes faster as $\mu$ approaches to $0.07$.
Note that the parameter setting $\mu > 0.9$ results in unstable behavior, i.e.,  
the ASE is diverging in some cases. This experimental results provides a justification of the 
use of (\ref{mu_rule}).

We then discuss a mismatched cooperative LMS with fixed $\mu$ determined by (\ref{mu_rule}).
The experimental conditions are exactly the same as the previous one. The only difference is that 
$\eta \in \{0.02, 0.04, 0.06, 0.08\}$ is used with fixed $\mu$.
Figure \ref{mismatched} (right) displays the ASE of the mismatched LMS with fixed $\mu$.
We can see that convergence becomes faster as $\eta$ grows but almost no improvement 
can be obtained for $\eta > 0.06$.  In this case, the average value of $\eta$ defined by (\ref{eta_rule})
is $0.05238$. This result implies that the value obtained by (\ref{eta_rule}) seems near optimal with respect to $\eta$.

In summary, from this experimental results, we can say that the parameter setting based on (\ref{eta_rule}) and (\ref{mu_rule}) are reasonable one to get sufficiently fast convergence of the cooperative LMS.

\begin{figure}
\begin{center}
\includegraphics[scale=0.4]{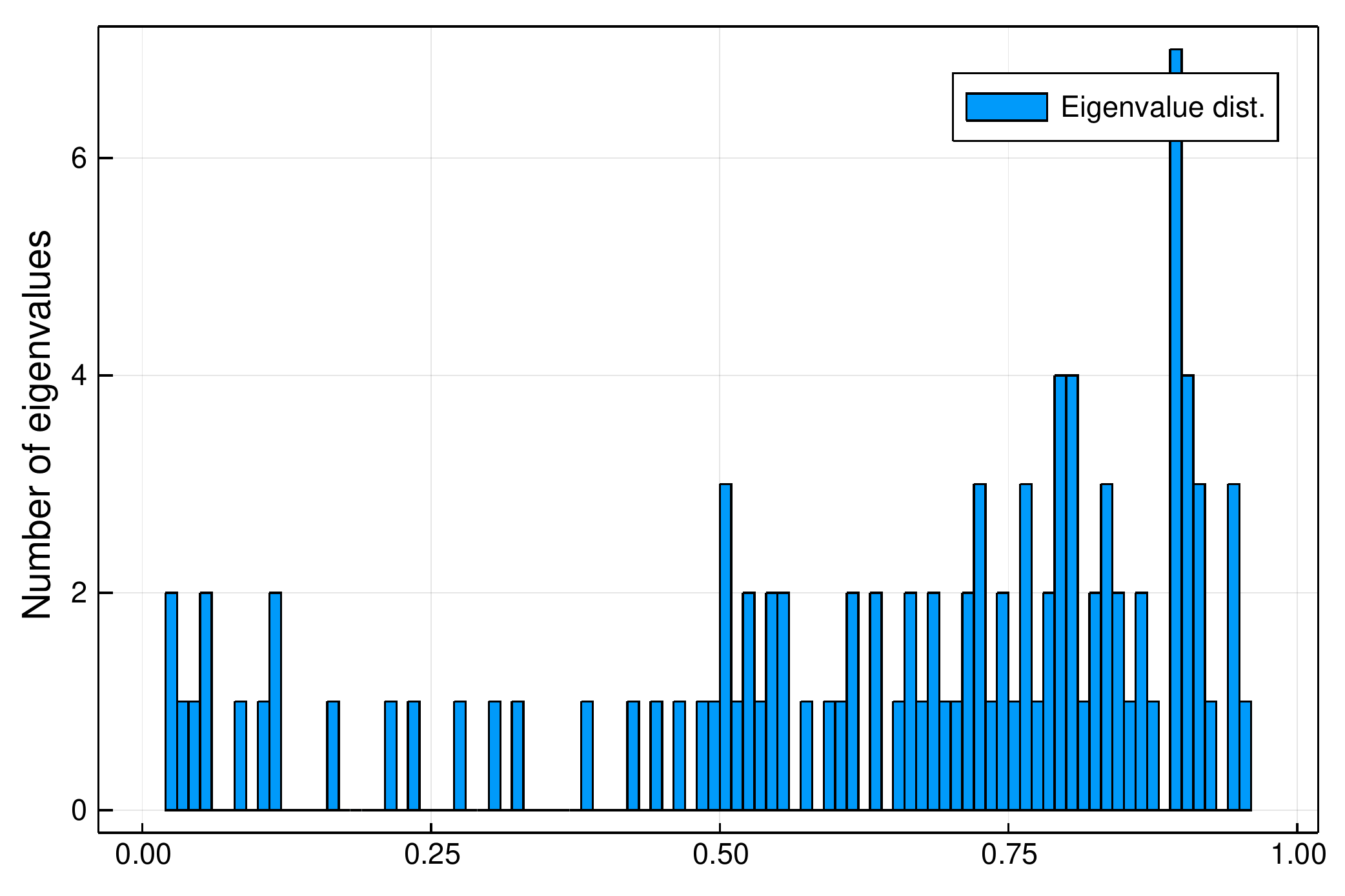}
\end{center}
\caption{Eigenvalue distribution of $\bm Q$ for Karate graph}
\label{karate-eigen}
\end{figure}

Figure \ref{karate-eigen} shows an eigenvalue distribution of the matrix $\bm Q$ under the 
same parameter setting.
We can confirm that all the eigenvalues of $\bm Q$ are included in the range $(0, 1)$ 
which is consistent with the claim of Theorem \ref{eigen_range}.
The parameters $\eta$ and $\mu$ are set to $\eta = 0.0523, \mu = 0.0688$ 
according to (\ref{eta_rule}) and (\ref{mu_rule}) in this case.

\subsection{Performance measure}

After the cooperative LMS is executed,  agent $k$ has its
own estimate $\bm{x}_k^{(L)}$. In this paper, the quality of the estimation 
is evaluated by the averaged squared error defined by
\begin{align}
	E_L := (1/K) \sum_{k=1}^K\|\bm{x}^* - \bm{x}_k^{(L)}\|^2,	
\end{align}
where $\bm x^*$ is the original LMS solution. If $\bm H_k (k \in [K])$ is stochastic, 
that is, randomly generated, the expectation $\mathbb{E}[E_L]$ will be estimated 
as the primary performance measure.
For the benchmark purpose, noncooperative LMS for each agent, that is, 
$
	\bm x^*_k := \text{argmin}_{\bm x \in \mathbb{R}^N}  \frac 1 2 \|\bm y_k -  \bm H_k\bm x \|^2,\  k \in [K]
$
will be considered. In such a case, the average LMS for noncooperative LMS is given by
$
	\tilde E_L = (1/K) \sum_{k=1}^K\|\bm{x}^*_k - \bm{x}_k^{(L)}\|^2.
$

\section{Chebyshev PSOR for Cooperative LMS}

It is natural to introducing Chebyshev-PSOR method  to the cooperative LMS 
for accelerating the convergence speed. In this section, we will show 
Chebyshev-accelerated cooperative LMS.

\subsection{Chebyshev interval}
The error evolution in (\ref{error_evolution}) can be rewritten as
$
\bm e^{(t+1)}  = \bm Q \bm e^{(t)}
$
where $\bm e^{(t)} := \bm\chi^{(t)} - \bm\chi^{*}$.  It can be seen as a linear fixed-point iteration.
When we apply PSOR to the cooperative LMS, the evolution can be described as
\begin{align} 
\bm e^{((\ell+1)T)} &= \left[\prod_{k = 0}^{T-1}(\bm I - \omega_k \bm B ) \right] \bm e^{(\ell T)},
\end{align}
where $\bm B := \bm I - \bm Q$. According to the strategy proposed in \cite{takabe20}, 
we can introduce Chebyshev PSOR factors for bounding the absolute values of eigenvalues of 
$\prod_{k = 0}^{T-1}(\bm I - \omega_k \bm B )$.
To obtain the best convergence rate, \cite{takabe20} proposed settings of
$a = \lambda_{min}(\bm  I - \bm Q)$ and $b = \lambda_{max}(\bm  I - \bm Q)$.
 However, in this paper, we will treat $a$ and $b$ as hyperparameters that define
the {\em Chebyshev interval} $[a, b]$.
This is because the discussion in \cite{takabe20} mainly considers the minimization of 
the spectral radius of $\prod_{k = 0}^{T-1}(\bm I - \omega_k \bm B)$ but 
the contributions of other eigenvalues to the error magnitude are not negligible 
when the number of iterations is small.
Namely,  for the cooperative LMS, an appropriate choice of the Chebyshev interval $[a, b]$
is important for obtaining a smaller error magnitude when the number of iterations is relatively small.

\subsection{Chebyshev cooperative LMS algorithm}

Algorithm \ref{cheb_alg} shows the details of the Chebyshev cooperative LMS Algorithm.
Lines 6-8 of the Chebyshev cooperative LMS algorithm represents corresponds to 
the additional PSOR process. The process in Line 8, 
\begin{align}
\bm x_k^{(t+1)} = (1 - \omega)\bm x_k^{(t)} + \omega \bm v_k^{(t)} 	
\end{align}
combines the input with the gradient step and the output from the proximal step.
This process can be carried out in each node in parallel.
Because the Chebyshev-PSOR factor $\omega$ should be the same for every agent, 
the loop index $t$ is needed to be synchronized for all the agents.
The additional computational complexity in Lines 6-8
is  negligible compared with the required 
computational complexity required to execute Lines 4 and 5.
\begin{algorithm}
 \caption{Chebyshev cooperative LMS}
 \label{cheb_alg}
 \begin{algorithmic}[1]
 \renewcommand{\algorithmicrequire}{\textbf{Input:}}
 \renewcommand{\algorithmicensure}{\textbf{Output:}}
%\REQUIRE in: $l_{min}, l_{max}$
% \ENSURE  out
  \STATE For each $k \in [K]$, set  $\bm x^{(0)} := \bm 0$
  \FOR {$t := 0$ to $L-1$}
  \FOR {$k := 1$ to $K$}
  \STATE \bm $\bm u_k^{(t)} 
  :=\bm x_k^{(t)} + \mu \bm H_k^T(\bm y_k - \bm H_k \bm x_k^{(t)})$ 
	\STATE $\bm v_k^{(t)} := \bm u_k^{(t)} + \eta \sum_{j \in {\cal N}(k)} (\bm u_j^{(t)} - \bm u_k^{(t)})$
	\STATE $t' := t \text{ mod } T$
	\STATE $\omega := \left[(b+a)/2 + ((b-a)/2) \cos(\pi (2t'+1)/(2T)) \right]^{-1}$
	\STATE $\bm x_k^{(t+1)} := (1 - \omega)\bm x_k^{(t)} + \omega \bm v_k^{(t)} $
	\ENDFOR
  \ENDFOR
 \RETURN $\bm x_1^{(L)}, \bm x_2^{(L)}, \ldots, \bm x_K^{(L)}$
 \end{algorithmic} 
 \end{algorithm}

\section{Experimental Results}
In this section, we will study empirical performance of the Chebyshev LMS.

\subsection{ASE performance of cooperative LMS for Karate graph}

\begin{figure}
\begin{center}
\includegraphics[scale=0.4]{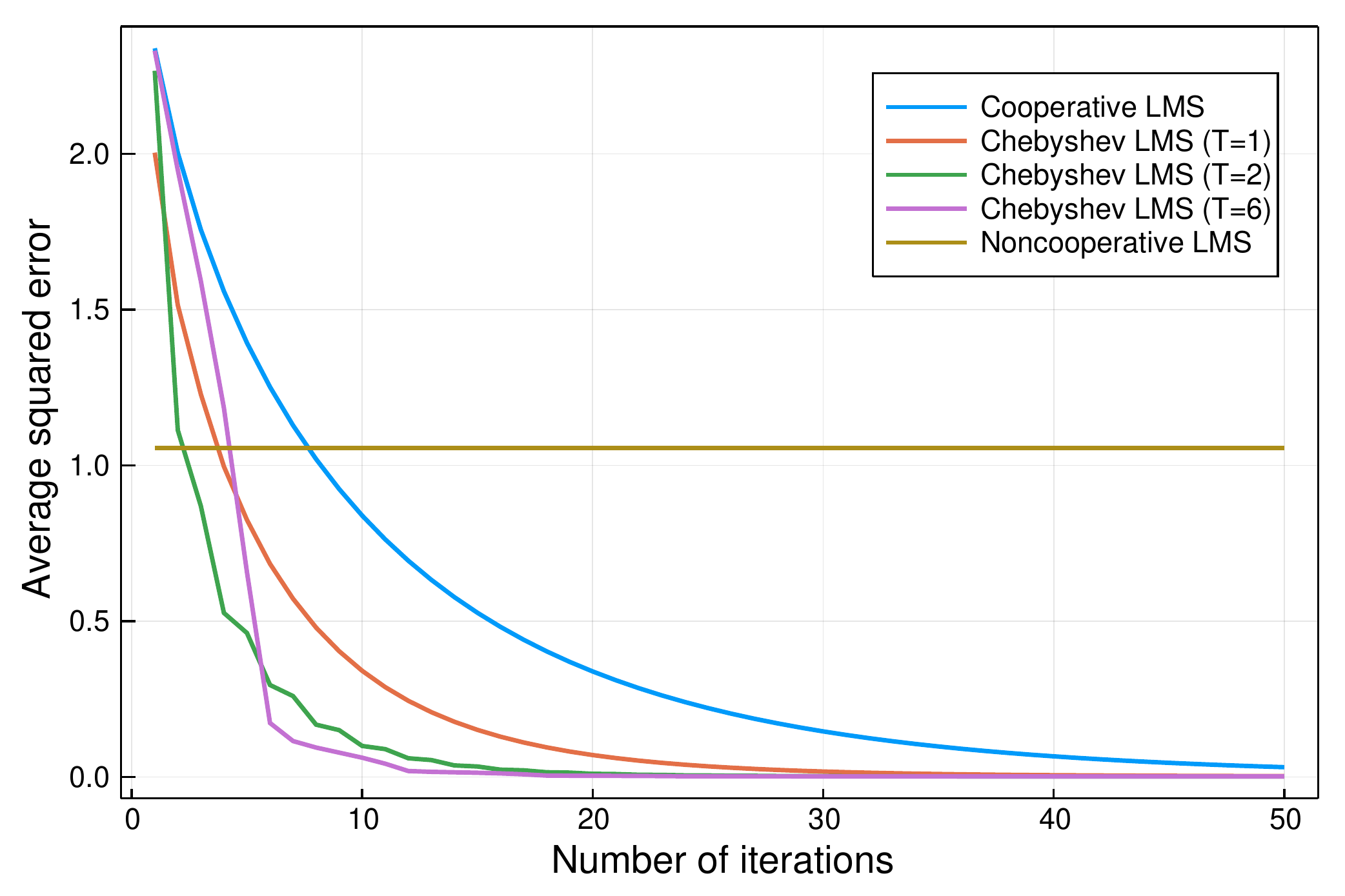}
\end{center}
\caption{Average squared errors for cooperative LMS for Karate graph (averaged over 100 trials) }
\label{karate_ase}
\end{figure}

The experimental conditions are summarized as follows.
We used Karate graph with $K = 34$.
The dimension of the $\bm x_0$ was set to $N = 3$, 
where each element in $\bm x_0$ followed ${\cal N}(0,1)$.
Each element in $\bm H_k $ also followed ${\cal N}(0,1)$ where 
$\bm H_k \in \mathbb{R}^{2 \times 3} (m = 2) $.
The standard deviation of the observation noises was set to $\sigma = 0.1$.
%The parameter $\epsilon$ used in (\ref{eta_rule}) and (\ref{mu_rule}) is set to $0.05$.
%In order to estimate the expectation, we run 100 trials.
Figure  \ref{karate_ase} presents the average squared errors (ASE)
of the original cooperative LMS defined in Alg.\ref{dist_LMS}
and the Chabyshev cooperative LMS defined in Alg.\ref{cheb_alg}
with $T = 1,2, 6$. 	  The Chabyshev cooperative LMS employed  the parameters $a = 0.15,	b = 1.0$,
and $\epsilon = 0.05$.
The expectation regarding the randomness of $\bm x_0$ and $H_k$ was estimated 
with 100 trials; that is,  we executed 100 runs of cooperative LMS with 
randomly initialized $\bm x_0$ and $\bm H_k$.
For comparison with the baseline performance, Fig. \ref{karate_ase} also shows 
the ASE for noncooperative LMS.

From Fig.  \ref{karate_ase}, it is immediately recognized that every cooperative LMS provides 
a decreasing ASE as the number of iterations increases.  For example, the original 
cooperative LMS provides a smaller ASE than noncooperative ASE after 10 iterations. 
This means that the advantage of the cooperative estimation appears after 10 iterations.
The Chebyshev LMS shows much faster convergence compared with the original 
cooperative LMS. When $T = 6$,  Chebyshev LMS requires only 10 iterations to achieve 
the ASE, while the original cooperative LMS needs 50 iterations to reach the same ASE.

\begin{figure}
\begin{center}
\includegraphics[scale=0.4]{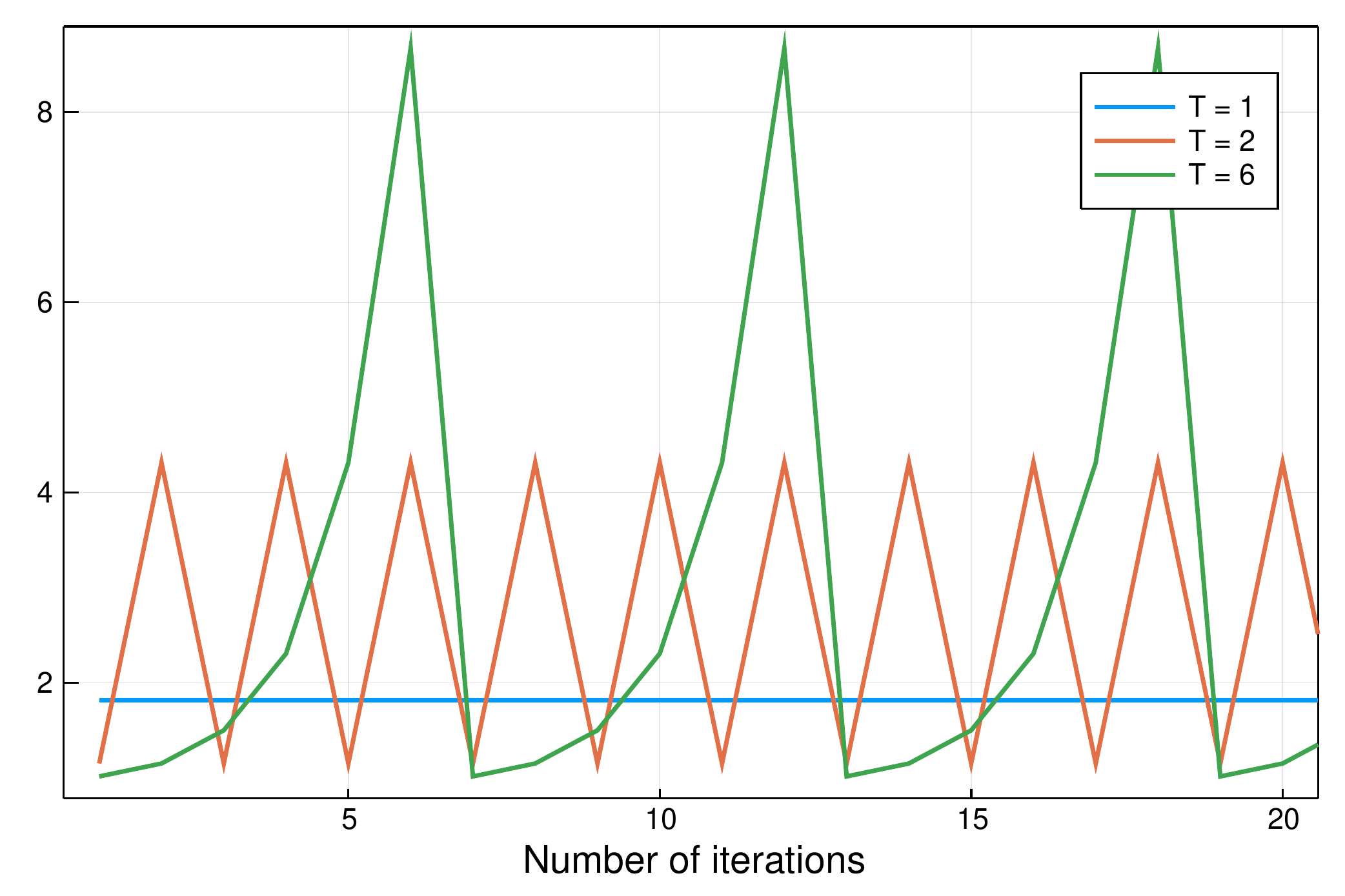}
\end{center}
\caption{Chebyshev factor $\omega$ as a function of the number of iterations for $T = 1,2,6$.}
\label{cstep}
\end{figure}

Figure \ref{cstep} depicts the Chebyshev factor $\omega$ as a function of the number of 
iterations. These factors are calculated under the assumption that $a = 0.1$ and $b = 1.0$.
We can see the zig-zag behavior of the Chebyshev factor.

%Figure \ref{normalized} presents normalized absolute eigenvalues of  
%$\prod_{k = 0}^{T-1}(\bm I_{KN} - \omega_k \bm B^* )$.  The horizontal axis represents 
%the value of eigenvalue of $\bm I_{KN} - \bm Q$. Let $\lambda$ be such an eigenvalue.
%The vertical axis represents the values of $|\prod_{k = 0}^{T-1}(\bm I_{KN} - \omega_k \lambda )|^{1/T}$
%where $(\cdot)^{1/T}$ is taken for fair comparison. 
%We can see that some eigenvalues takes high values around $\lambda = 0$ but 
%the absolute values are well suppressed in the remaining range when $T = 2, 4$
%\begin{figure}
%\begin{center}
%\includegraphics[scale=0.4]{../Experiments/Karate/norm.pdf}
%\end{center}
%\caption{Normalized absolute eigenvalues of  $\prod_{k = 0}^{T-1}(\bm I_{KN} - \omega_k \bm B^* )$ for $T = 1,2,4$. 
%The hyper parameter setting is: $a = 0.15, b=1.0$.}
%\label{normalized}
%\end{figure}

\subsection{Small graphs}

\begin{figure}
\begin{center}
\includegraphics[scale=0.4]{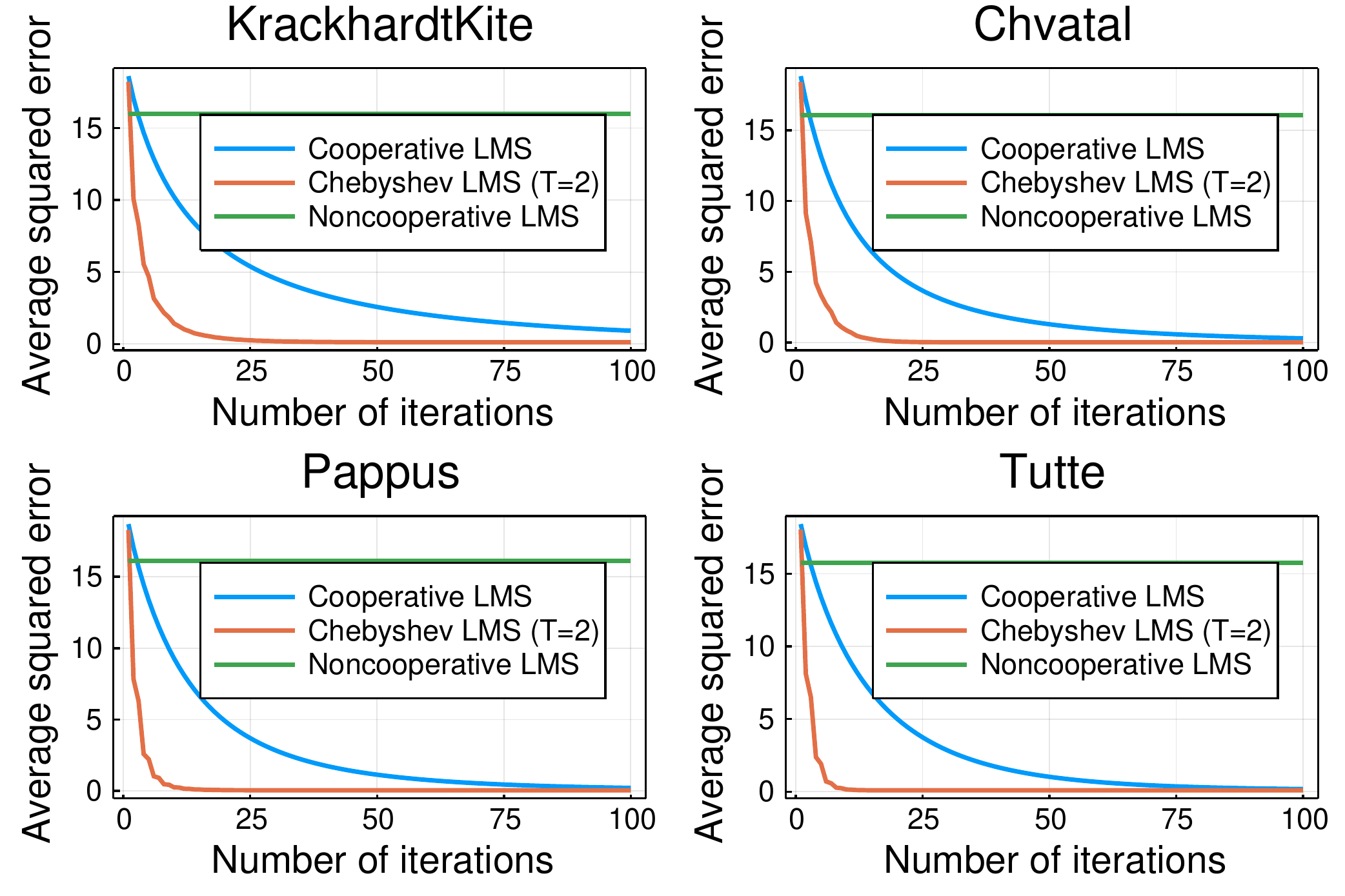}
\end{center}
\caption{Average squared errors for cooperative LMS for small graphs}
\label{smallgraphs}
\end{figure}

In the previous subsection, we examined the performances of the cooperative LMS for Karate graph.
We here study the performance of the cooperative LMS for several well-known small graphs,
such as Krackhardt kite, Chv\'atal, Pappus, and Tutte.
The number nodes and edges in these graphs are summarized in Table~\ref{smallgraph_table}.
The main parameters were
$N = 20, m = 5, \sigma = 1.0$, $a = 0.15,	b = 1.0, \epsilon = 0.05$.
\begin{table}
\caption{Small graphs treated in this subsection}
\label{smallgraph_table}
\begin{center}
\begin{tabular}{lcc}
\hline
 & $|V|$ & $|E|$ \\
\hline
Krackhardt kite & 10 & 18 \\
Chv\'atal & 12 & 24 \\
Pappus & 18 & 27 \\
Tutte & 46 & 69 \\
\hline
\end{tabular}	
\end{center}
\end{table}
Figure \ref{smallgraphs} summarizes the ASE performance for these small graphs.
We can see that Chebyshev LMS achieves much faster convergence for all the graphs.
 
 \subsection{Random graphs}
 
 We here examine the ASE performance of the Chebyshev LMS for 
 moderately large random graphs. We selected two types of
 random ensembles,   Erd\"os-R\'enyi (ER) and Barab\'asi-Albert (BA) random graphs.
 
 \subsubsection{ER random graph}

\begin{figure}
\begin{center}
\includegraphics[scale=0.4]{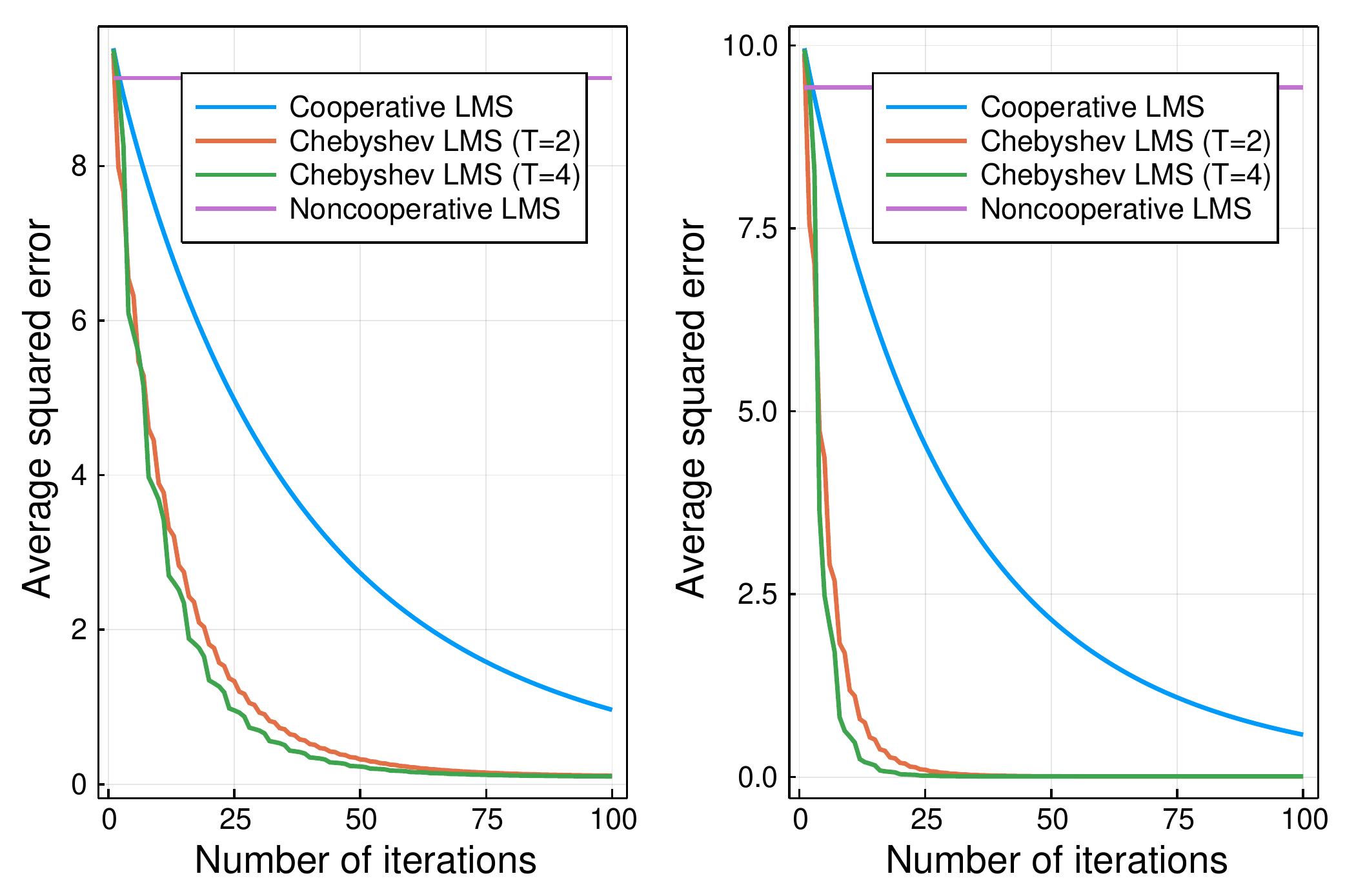}
\end{center}
\caption{Average squared errors for cooperative LMS for ER graphs of size $K = 100$: (left) sparse ER graph with $p = 0.05$, (right) 
dense ER graph with $p = 0.25$}
\label{ERgraphs}
\end{figure}

We first examine ER random graph ensembles.
The main parameters were
 $N = 10, m = 1, \sigma = 1.0$, $a = 0.2,	b = 1.0, \epsilon = 0.05$.
Figure \ref{ERgraphs} shows the ASE of cooperative LMS for 
sparse $(p = 0.05)$ and dense ($p = 0.25$) cases.
In both cases, we can observe that Chebyshev LMS provides faster convergence.
Furthermore, for the dense graphs (right), the convergence of ASE of Chebyshev LMS
is faster than that for the sparse graphs.

\subsubsection{BA random graph}

\begin{figure}
\begin{center}
\includegraphics[scale=0.4]{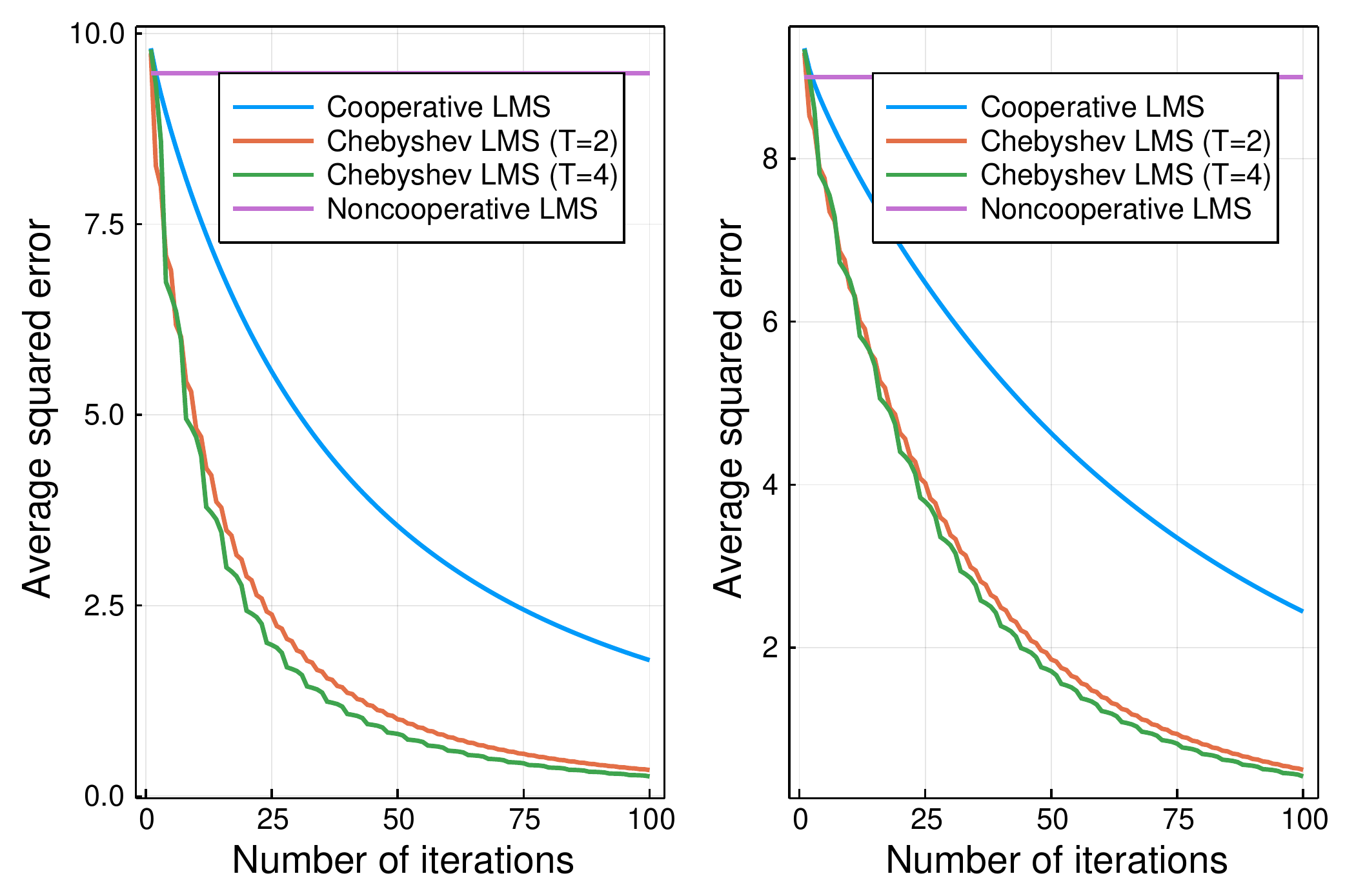}
\end{center}
\caption{Average squared errors for cooperative LMS for BA graphs (left) small BA graph with $K = 30$, (right) 
large BA graph with $K = 200$}
\label{BAgraphs}
\end{figure}

As an example of random scale-free networks, 
we here evaluate BA random graphs that use a preferential attachment mechanism. 
The number of edges between a new node and  existing nodes was set to 3.
The main parameters were
 $N = 10, m = 1, \sigma = 1.0$, $a = 0.2,	b = 1.0$ and $\epsilon = 0.05$.
Figure \ref{BAgraphs} shows the ASE of cooperative LMS for 
small $(K=30)$ and large ($K=200$) cases.
It is immediately observed that small networks (left) show faster convergence 
than larger networks (right). This seems a natural consequence 
of the required propagation time, that is, message propagation needs more time 
for a large graph. Although the convergence becomes slower,
Chebyshev LMS shows reasonable acceleration even for a large graph. 

\section{Concluding Summary}

In this paper,  we described how Chebyshev PSOR can be successfully applied 
to a distributed LMS algorithm. Accelerations of convergence speed has been 
empirically confirmed in many distributed LMS scenarios. 
The cooperative LMS algorithm presented in this paper 
includes two step size parameters $\eta$ and $\mu$ which 
are assumed to be shared with every agent.
This means that we need a method to share $\eta$ and $\mu$ in a distributed manner.
The largest eigenvalue of 
the graph Laplacian $\bm L$ can be evaluated by a 
decentralized  method \cite{Aragues}. If the graph topology remains the same,
we only need the initial computation of the largest eigenvalue. 
On the other hand, the largest eigenvalues 
of the gram matrix $\bm H_k^T \bm H_k$ can be estimated by using
Marcenko-Pastur law or an empirical upper bound based on the statistics of $\bm H_k$.
Combining these pre-parameter sharing processes, the proposed Chebyshev LMS algorithm 
becomes fully distributed algorithm.

It is highly expected that the principle shown in this paper has
straightforward applicability to other distributed signal processing algorithms
if the node operation can be described as an affine or linear transformation.
When the node operation contains non-linear mapping, the methodology 
presented in the paper may not be directly exploited. 
Extension towards such a situation is an interesting open problem.

\end{document}